\documentclass[useAMS,usenatbib,fleqn]{mn2e}
\usepackage{amsmath,amssymb,amsfonts,latexsym}
\usepackage[]{graphicx}
\usepackage{subcaption}
\usepackage{longtable}
\usepackage{fixltx2e}
\usepackage{natbib}
\usepackage{hyperref}

\newcommand{\be}{\begin{equation}}
\newcommand{\ee}{\end{equation}}

\newcommand{\rs}{\rho_{s}}

\def\be{\begin{equation}} 
\def\ee{\end{equation}} 
\def\beq{\begin{eqnarray}} 
\def\eeq{\end{eqnarray}}

\def\04a{{2004 a}}
\def\04b{{2004 b}}

\title{Heating of the Real Polar Cap of Radio Pulsars }

\author[M.~Sznajder\& U.~Geppert] {M.~Sznajder$^{1}$\thanks{E-mail:Maciej.Sznajder@dlr.de}, U.~Geppert$^{2}$\\
$^1$ German Aerospace Center, Institute of Space Systems, Robert-Hooke-Str. 7, 28359 Bremen, Germany\\
$^2$ J. Gil Institute of Astronomy, University of Zielona G\'{o}ra, ul. Szafrana 2, 65-516, Zielona G\'{o}ra, Poland }

\begin{document}

\date{}

\maketitle

\label{firstpage}

\begin{abstract}
The heating of the real polar cap surface of radio pulsars by the bombardment of ultra-relativistic charges is studied. The real polar cap is a significantly smaller area within or close by the conventional polar cap which is encircled by the last open field lines of the dipolar field $\vec{B}_d$. It is surrounded by those field lines of the small scale local surface field $\vec{B}_s$ that join the last open field lines of $\vec{B}_d$ in a height of $\sim 10^5$ cm above the cap. As the ratio of radii of the conventional and real polar cap  $R_{dip}/R_{pc}\sim 10$, flux conservation requires $B_s/B_d\sim 100$. For rotational periods $P\sim 0.5$ s, $B_s\sim 10^{14}$ G creates a strong electric potential gap that forms the inner accelerating region (IAR) in which charges gain kinetic energies $\sim 3\times 10^{14}$ eV. This sets an upper limit for the energy that back flowing charges can release as heat in the surface layers of the real polar cap. Within the IAR, which is flown through with a dense stream of extremely energetic charges, no stable atmosphere of hydrogen can survive. Therefore, we consider the polar cap as a solidified ``naked'' surface consisting of fully ionized iron ions. We discuss the physical situation at the real polar cap, calculate its surface temperatures $T_s$ as functions of $B_s$ and $P$, and compare the results with X-ray observations of radio pulsars.  \end{abstract}

\begin{keywords}
stars: neutron - stars: magnetic fields - pulsars: general - stars: interiors
\end{keywords}

\section{Introduction}
\begin{figure}
\centering
\includegraphics[width=1.0\columnwidth]{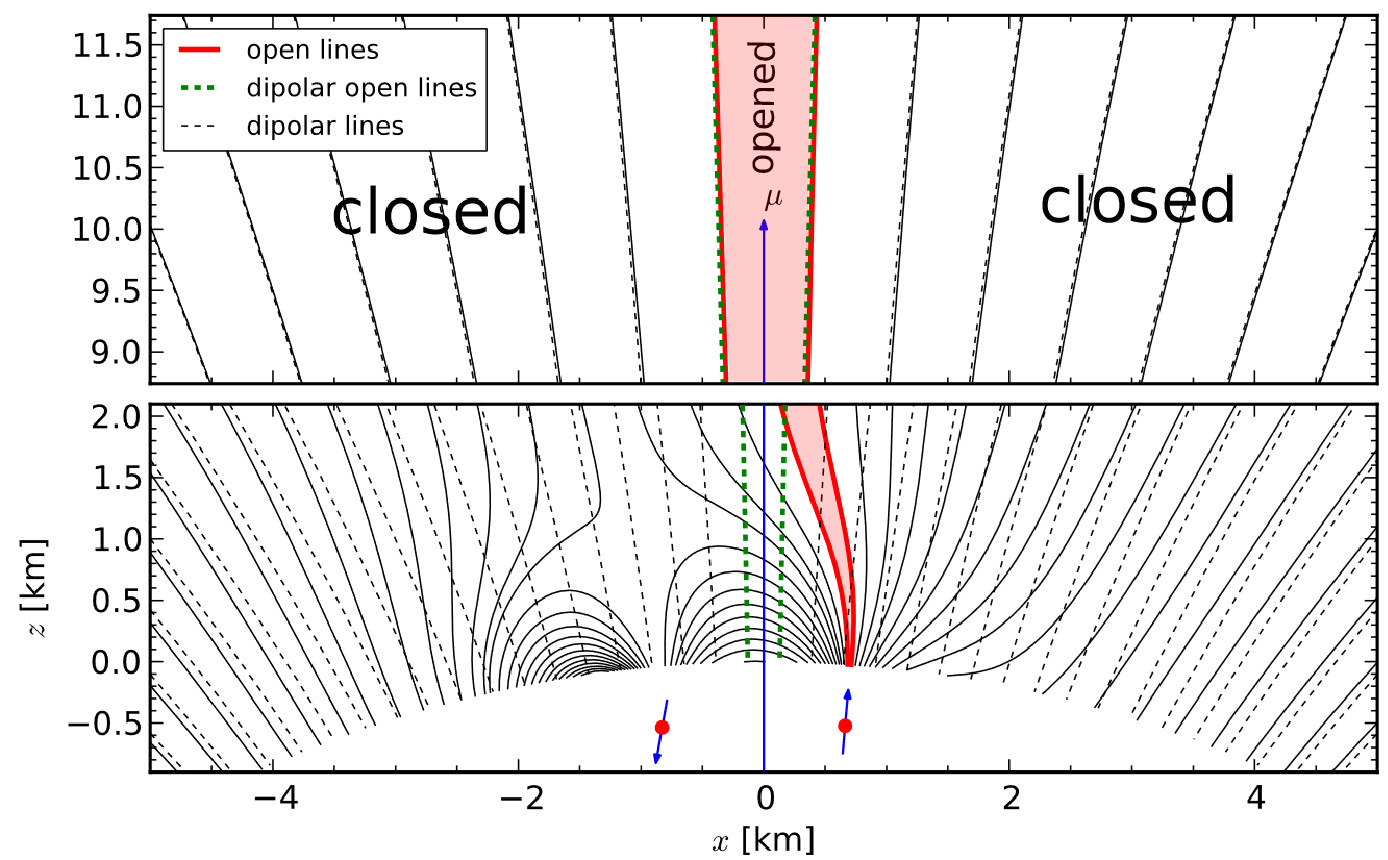}
\caption{Schematic presentation of the local magnetic field structure at and above the polar cap. The figure is taken from \citealt{SMG15b}. $z$ denotes the height above the polar cap surface, $x$ the distance from the pole, $\mu$ is the magnetic dipole moment; the dashed lines represent the field lines of the large scale dipolar field while the full lines indicate those of the local small scale field structure.}
\label{fig:field_structure}
\end{figure}

Pulsars function as radio emitters if the complicated interplay between sufficiently fast rotation, structure, and strength of the magnetic field at the polar cap surface $\vec{B}_s$, and the local surface temperature $T_s$ are efficiently coupled. Both in the vacuum gap model \citep{RS75} and in the space charge limited flow model \citep{AS79} $\vec{B}_s$ has to have a scale of $\lesssim 10^6$ cm, significantly smaller than the scale of the dipolar field $B_d$ that determines the braking of the pulsar rotation.\\
\noindent In the vacuum gap model and its refinements \citep{GMG03}, a sufficiently large cohesive energy of the charges in the surface layer of the polar cap is necessary to allow the maintenance of an electric potential gap $\Delta V$ above it. The cohesive energy increases with $B_s$ but decreases with increasing $T_s$. \cite{ML07} have shown that in an iron cap surface with $T_s\sim 10^6$ K a gap can be formed only if $B_s\gtrsim 5\times 10^{13}$ G.\\
\noindent  There are observational evidences and theoretical arguments that indicate the existence of strong and small scale magnetic field components in the polar cap region. They are characterized by curvature radii $R_{cur}\lesssim 10^6$ cm instead of $\sim 10^8$ cm as expected for $\vec{B}_d$. The presence of small scale field structures at the surface of neutron stars has been studied by \cite{IEP16} who discussed both observational and theoretical arguments that these field structures can survive the fall-back episode, and be re-established by the Hall drift and maintained for $\gtrsim 10^6$ yr; a result obtained also by \cite{GV14}.  \cite{LG19} studied the failing of a neutron star's crust which can be understood only by the presence of small scale surface fields with $B_s > 10^{15}$ G. Also, the misalignment between the thermal X-ray and the radio emission peak indicates the presence of multipolar field components at the inner accelerating region (IAR) above the polar cap \citep{AM19,PM20}.
Recently, \cite{LGOP19} found in the X-ray light curves of the millisecond pulsar J0437-4715 evidences for the presence of small-scale field structures at the polar cap. Similar structures have been seen by \cite{LCPRR19} in the X-ray light curves of the magnetar J1745-2900. Therefore, the existence of such surface field structures is a widely observed phenomenon on neutron stars. \\
\noindent The polar cap surface is heated by its bombardment with backflowing charges that acquired ultra-relativistic energies within the inner accelerating region (IAR) above the ``real'' polar cap. Simple estimates from the balance of heat power density input  $e\Delta{V_{max}}c n_{GJ}$ ($n_{GJ}$ is the Goldreich-Julian charge density) and the blackbody power density $\sigma_{SB} {T_s}^4$ indicate a typical $T_s\sim 10^6$ K.\\
\noindent What defines the ``real'' polar cap? The canonical radius of the polar cap area $R_{dip}$  is given by the last open dipolar field line and depends on the light cylinder radius, i.e. on the pulsar's rotational period \citep{RS75}.\\
\noindent A few years ago, simultaneous radio and X-ray observations were performed revealing an area bombarded by the backflow of relativistic charges, i.e. the part of the cap surface where the strong and small scale field is anchored and which is responsible for the pair creation. This area is significantly smaller than the conventional polar cap area, \citep{H13, SGZHMGMX17} and has a temperature $T_s \sim 2\ldots 5\times 10^6$ K. Blackbody fits of thermal X-ray spectra provide radii $R_{pc}$ of the emitting area, which is smaller by a factor $\sim 10$ than the radius of the conventional polar cap $R_{dip}$ (see Table 1 in \cite{G17}). Flux conservation arguments indicate that if $R_{dip}/R_{pc} \sim 10$, $B_s\sim 100 B_{dip}$ so that for a typical $B_{dip}\sim 10^{12}$ G $B_s$ at the real polar cap is in the order of $10^{14}$ G. Such field strengths cause a sufficiently high cohesive energy, in the cap surface layers, necessary for the creation of an electric potential gap sufficiently high to guarantee copious pair production. This small hot and highly magnetized bottom area of the IAR will hereafter referred to as real polar cap. Its $\vec{B}_s$ has almost poloidal magnetic field structures and joins the global dipolar field in at a height of a few $\sim 10^5$ cm, roughly the curvature radius of $B_s$. A representation of the magnetic field structure in the region of the polar cap is presented in Fig.~\ref{fig:field_structure}.\\
\noindent Clearly, the blackbody fits which determine the radius of the real polar cap $R_{pc}$ and $T_s$ should be considered with caution. Due to the variance in the photon statistical data, both thermal and non-thermal X-ray spectra can be fitted equally well to various models \citep{AM19}. However, the availability of $B_d, B_s, R_{dip}, R_{pc},$ and $T_s$ for a number of radio pulsars is reason enough, albeit with caution, given the large error bars at $T_s, R_{pc},$ and $B_s$,  to study the establishment of the surface temperature of the real polar cap in greater detail.\\
\noindent Recently, two studies on the polar cap heating by return currents of radio pulsars have been published \citep{T17, BPO19}. A basic aspect of such studies is the energy of the electrons/positrons that hit the polar cap surface. This energy is determined by the strength and structure of the local surface magnetic field at the cap $\vec{B}_s$, by the rotational period $P$ of the pulsar, and by the height $h$ of the IAR. Therefore, these parameters define the strength of the electric field that is eventually capable of accelerating charged particles. For fundamentals, see \cite{RS75} and \cite{GMG03}.\\
\noindent \cite{T17} considers a typical situation with $B_s \sim 10^{12}$ G and $P\sim 1$ s. In this model, return currents heat up a small semi-ring area along the rim of the polar cap. Besides the fact that $B_s \sim 10^{12}$ G is too weak to allow a potential gap formation, sites of heat release at the bottom of the IAR are the sparks, arranged in equidistant patterns over the whole polar cap surface (see Fig. 1 of \cite{GS00}). Therefore, the whole cap, not only the rim region, is at $T_s$. If the heat would be released only around the rim, the strong almost poloidal $\vec{B}_s$ would prevent the heat flux from the rim moving onto the whole polar cap surface. \cite{TA13} and \cite{BS11} have shown that for inclination angles between dipole and rotation axis, $\chi \geq 60^\circ$ super- or anti- Goldreich-Julian current regions appear within the conventional polar cap. For the oblique rotator case, these currents fill the whole cap area. These currents will enhance the pair creation, thereby enforcing the heating of the polar cap surface.\\
\noindent \cite{BPO19} assume the existence of an atmosphere above the polar cap. While the larger area of the neutron star surface may be covered by an atmosphere, it is unlikely that within the IAR, a small cylinder of $\sim 5\cdot 10^3$ cm height with a diameter of $\sim 10^3\ldots 10^4$ cm, where extremely energetic charged particles pass through, can sustain a stable atmosphere of hydrogen.
\noindent Therefore, it is plausible that  neither the assumption of $B_s\sim 10^{12}$G nor the existence of an hydrogen atmosphere in the IAR, nor the Ohmic dissipation of return currents solely in the polar cap rim region are realistic expectations explaining $T_s\sim 2\ldots 5\times 10^6$ K of the real polar cap.\\
\noindent The aim of this study is to prove, whether under conditions present at and above the real polar cap, if the surface temperature can be held at a few $10^6$ K over the typical lifetime of radio pulsars. Herein is considered the bombardment of the cap surface by ultra-relativistic electrons, i.e. pulsars with $\vec{\Omega} \cdot \vec{B} <0$, $\Omega$ being the pulsar's rotational frequency.  In Section 2, is the discussion of the physical situation at the real polar cap and in the IAR. Section 3 is devoted to estimate the kinetic energies, and the flux of the electrons as functions of $B_s$ and $P$. By solving the Bethe-Bloch equation, the penetration depths of the electrons, their heat release, and surface temperatures as functions of the electron kinetic energies are presented. In Section 4, comparison of the obtained results with observations of $T_s$ are made, and concluded in Section 5.\\

\section{Physical situation at the polar cap of radio pulsars}
\begin{figure}
\centering
\includegraphics[width=1.0\columnwidth]{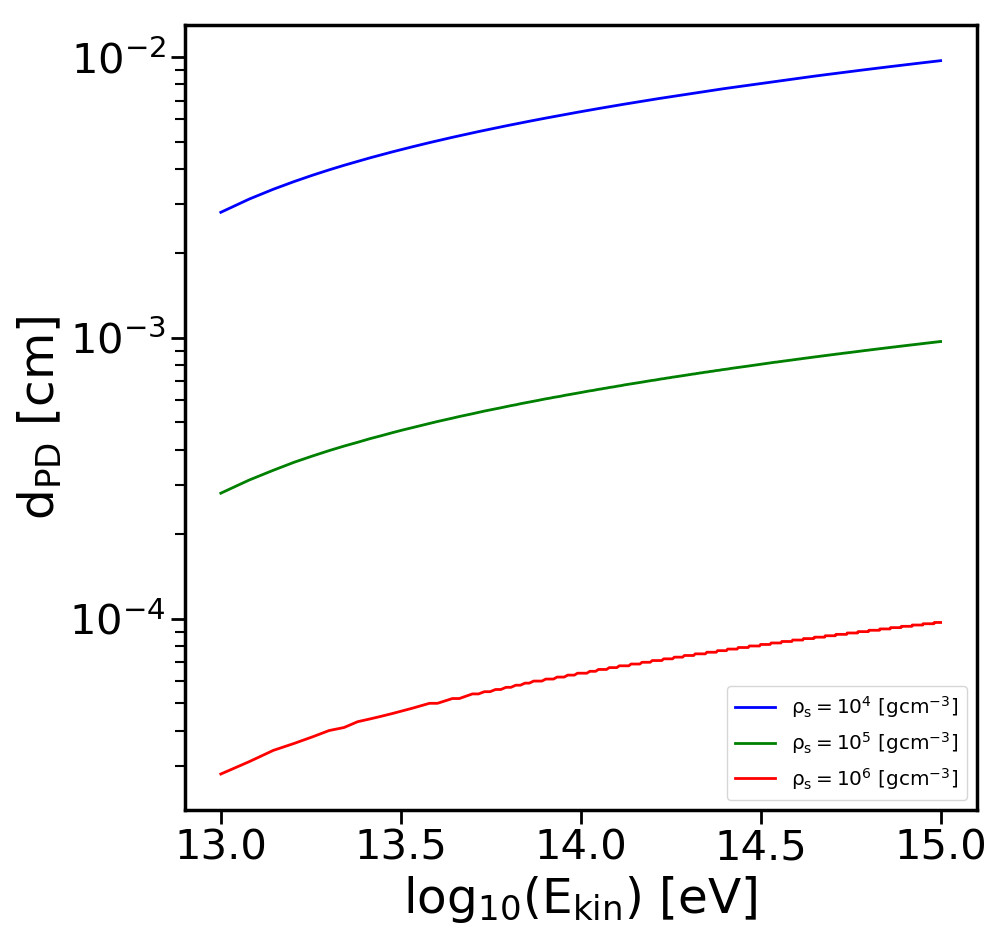}
\caption{Penetration depth of electrons into fully ionized iron for surface densities $\rho_s=10^4, 10^5, 10^6$ g cm$^{-3}$ }
\label{fig:pendepth}
\end{figure}

\noindent There is perhaps no place throughout a neutron star, where the magnetic, thermal, and rotational evolutions are more intensely coupled than at the real polar cap of radio pulsars.  Magneto - thermal processes at the polar cap operate on shorter timescales and within a smaller spatial region than in case of the magneto - thermal interactions, as recent thorough studies indicate; see e.g. \cite{VRPPAM13, PMG09, GV14, GMVP14}. It is widely accepted that pulsars create their radio emission by charges which are accelerated to ultra-relativistic energies either in a space charge limited flow of electrons and positrons, \citep{AS79} or in a vacuum gap \citep{RS75} just above the pulsar's polar cap.\\
As polar cap is considered conventionally as the area at the magnetic south and north poles of pulsars encircled by the last open field lines of the dipolar magnetic field $\vec{B}_d$ \citep{RS75}. As discussed in the Introduction, the physical situation of the real polar cap, the bottom of the IAR, \citep{GMG03} and its sub-surface layers are the current focus. The radius of the real polar cap is defined by the structure of $\vec{B}_s$ whose field lines join the open field lines of $B_d$ (see Fig.~\ref{fig:field_structure}). Here, the charges accelerated within the IAR reach regions close to the light cylinder where they eventually emit radio waves.\\
\noindent  Striking features are the large temperature differences between the polar cap and the rest of the neutron star surface. Meridional temperature gradients across the rim of the real polar cap are caused by the tremendous strength and almost radial direction of $\vec{B}_s$. A magnetization parameter $\omega_B\tau\gtrsim 100$ at a surface density $\rho_s\sim 10^6$ g cm$^{-3}$ \citep{G17} suppresses the meridional component of the heat conductivity resulting in the meridional heat flux being a factor of at least $10^4$ smaller relative to the radial flux \citep{HUY90, GKP04}. The rest surface cools according to well understood cooling scenarios (URCA or DURCA, photons), thereby increasing the meridional temperature gradient over time. After $\sim 1$ Myr the large part of the surface has a temperature in the order of a few $10^5$ K \citep{PGW06,VRPPAM13}.\\
\noindent These large temperature gradients, both in radial and meridional direction are restricted to a relatively shallow layer of the real polar cap beneath its surface. The kinetic energy of the backflowing charges is released as heat within a few radiation lengths, and is almost immediately re-radiated \citep{CR80} as practically no thermal heat is transported into deeper regions of the crust (see  Fig.~\ref{fig:pendepth}). The lifetime of radio pulsars is typically $10^6\ldots 10^7$ yr. Over this lifetime, the real polar cap area must be significantly hotter than the remaining surface of the neutron star.\\ 
\noindent How the electron - positron pairs are created and subsequently accelerated, and how they eventually cause the observed radio emission and the observed real polar cap heating, depends on the strength and structure of the local surface magnetic field $\vec{B}_s$, on the rotational velocity, and on the angle between the axis of rotation and magnetic field. Already \cite{RS75} and \cite{AS79} noted that for a sufficiently powerful creation of electron-positron pairs, the magnetic field at the surface of the polar cap must be significantly more curved than the far above the neutron star surface dominating dipolar field $\vec{B}_d$. The latter has a curvature radius of $\sim 10^8$cm, while copious pair production requires curvature radii $\lesssim 10^6$cm. A mechanism that creates the strong and small scale field structures, at the polar cap surface, could be the crustal Hall drift \citep{RBPAL07, PG07, GH18}. It may create the required short scale poloidal field structure out of a rich reservoir of magnetic energy stored in a toroidal field located deep in the crust and in the outer core layers. This would also enhance the surface temperature of the spot through local Ohmic dissipation \citep{GGM13,GV14}. However, this hot spot is larger than the real polar cap. The latter perhaps lies within the former, or shares some overlap. While the shape of the hot spot is formed by the crustal Hall drift, the surface of the real polar cap is determined by the small scale component of $\vec{B}_s$ which joins the open field lines of $\vec{B}_d$. While the surface temperature of the Hall drift hot spot never exceeds $1.5\times 10^6$ K (see Fig. 5 in \cite{GV14}), the surface temperatures of the real polar cap may even reach $\sim 5\times 10^6$ K as (though with large error bars) blackbody fits of B2224 +65 \citep{HHTTTWC12} or B1451 -68 \citep{PPMKG12} indicate.\\
\noindent An important subject is the surface density $\rho_s$ at the hot polar cap, where the neutron star matter is condensed; either liquid or solidified. The most recent description of the state of aggregation in that region is given by \cite{PC13}. The so-called zero pressure density can be considered as the surface which is hit by the bombardment with ultra-relativistic electrons/positrons. It increases with the local magnetic field strength  proportional to $B_{12}^{6/5}$, and can be estimated by
\beq
\rho_s=561\xi A Z^{-3/5}B_{12}^{6/5}\,\,, \xi \approx 0.517+0.24 B_{12}^{1/5}\, ,
\label{eq:rho_s}
\eeq

\noindent where $B_{12}=B/10^{12}$ G. For an Fe-surface ($A=56, Z=26$) is $\rho_s \approx 1.25\cdot 10^6$ g cm$^{-3}$  for local field strengths of $B_{12}=100$, as expected to be present at the polar cap surface.\\
Whether the surface matter is liquid or solid impacts the magnetic field dependent melting value of the Coulomb coupling parameter $\Gamma_m(B)$; if it exceeds $\sim 175$ the matter is solidified, below it is liquid. According to \cite{PC13} is 

\beq
\Gamma_m(B)\approx \Gamma(0)\left[1+0.2\beta\right]^{-1}\, .
\label{eq:GammaB}
\eeq

\noindent $\beta$ is given by the ratio of ion cyclotron to plasma frequency, $\beta\approx 0.0094 B_{12}\rho_6^{-1/2}$ with $\rho_6=\rho/10^6$ g cm$^{-3}$. $\Gamma(0)$ is roughly the ratio of Coulomb and thermal energy

\beq
\Gamma(0)=\frac{(Ze)^2}{a_ik_BT}\approx \frac{22.747}{ T_6}\rho_6Z^2A^{-1/3}\, ,
\label{Gamma0}
\eeq

\noindent with $a_i$ being the spacing between ions, and $k_B$ the Boltzmann constant. Inserting typical values for the surface temperature at the polar cap $T_6\approx 3\ldots 5$ (see Table 1 in \cite{G17}) and $\rho_{s,6}\approx  1.25$,  one finds $\Gamma_m \approx  1235$  for $T_6=3$ and $B_{12}= 100$. Thus, $\Gamma_m(B)$ is significantly greater then $175$. It is the process of magnetic condensation that appears for $B_s \gtrsim 10^{14}$ G \citep{LS97, TZD04, ML07, PC18}. Therefore, the polar cap surface of radio pulsars is probably in a solidified state. We will consider the real polar cap surface, consisting of fully ionized iron, as that of a ``naked'' neutron star \citep{TZD04} without any atmospheric layers above it.\\

\section{Heating of the polar cap by bombardment with ultra-relativistic electrons}
A basic feature of radio pulsar emission is the creation of electron-positron pairs in the IAR just above the real polar cap. In the case where the rotational axis and the magnetic field axis are anti-parallel ($\vec{\Omega}\cdot \vec{B}<0$), the positrons escape and eventually generate the radio emission while the electrons are accelerated toward the cap surface, heating it up to $1...5\times 10^6$ K \citep{RS75, AS79, GMG03}.\\
\noindent The amount of heat that the bombarding electrons release in the surface layers of the cap depends on the rotational period $P=2\pi/\Omega$, the local surface magnetic field strength $B_s$, and the height of the inner accelerating gap $h$ (see \cite{GMG03}, Eq. A.5  -A.9). Given these values the maximum potential drop in the inner accelerating gap is 

\beq
\Delta V_{max}=\frac{2\pi}{cP}B_s h^2\, .
\label{eq:Vmax}
\eeq
\noindent For typical parameters at the real polar cap, $B_s\sim 10^{14}$ G, $P=0.5$ s, $h=5\cdot 10^3$ cm, a maximum potential gap of $\sim 3\cdot 10^{14}$ V will be created. A shielding factor that takes into account the thermal detachment if iron ions from the cap surface may reduce $V_{max}$ by a factor of ten.

\noindent The kinetic energy $E_{kin}$ gained by an elementary charge $e$ that is accelerated within this gap  is given by

\beq
E_{kin}=e\Delta V_{max}\approx 3\cdot 10^{12}\frac{B_{s,12}h_{5,3}^2}{P_{0.5}}\, eV\, ,
\label{eq:eVmax}
\eeq

\noindent where $B_{s,12}$ is $B_s$ in $10^{12}$ G, $h_{5,3}$ the gap height in $5\cdot 10^3$ cm, and $P_{0.5}$ the rotational period in $0.5$ s. Since, the local surface magnetic field strength at the polar cap $B_{s,12}\gtrsim 100$, the kinetic energy of an elementary charge  $\approx 3\cdot 10^{14}$ eV. This kinetic energy of the primary particles corresponds to a Lorentz factor $\gamma \sim 6\cdot 10^8$. Such large $\gamma$ will be reached only at the beginning of each sparking cycle that lasts $\sim 10\mu\mathrm{s}$. Towards the end of a cycle the potential drop will decrease significantly, due to separation of the large electron/positron densities which have been produced in the discharge \citep{RS75, MGP00}, and by the thermal release of iron ions which cause an additional screening of the potential gap \citep{GMG03}. The secondary particle electron/positron plasma produced either by curvature radiation, or inverse Compton scattering within the IAR is more dense but less energetic. Its $\gamma$-factor is about 4 orders of magnitude smaller than that of the primary particles. Secondary charges produced outside the IAR don't contribute to the heating of the real polar cap surface.\\
\noindent The heating of the IAR bottom proceeds via the bombardment both with primary and secondary charges; the kinetic energy given by Eq.~\ref{eq:eVmax} is those of the primary particles.  How many secondary particles are produced by one primary charge and can hit the polar cap surface depends on the height $h$ of the IAR, and is determined by the multiplicity \citep{S71,TH15}. For $B\gtrsim 3\cdot 10^{12}$ G and $R_{cur}\sim 10^7$ cm \cite{TH19} find a maximum multiplicity  $\sim 10^6$. For stronger magnetic fields and smaller radii of curvature, no higher multiplicities will appear as photon absorption proceeds near the pair formation threshold. Thus, although the kinetic energy of the secondary charges is orders of magnitude smaller than that of the primary ones, this reduction will be counteracted by the high multiplicity. Half of the in-pairs created particles will leave the IAR. Since the kinetic energy of the primary particles is redistributed to the secondary ones, $E_{kin}$ of the primary particles gained in the IAR forms the kinetic energy budget available for the polar cap heating. Thus,  $E_{kin}$ given by Eq.~\ref{eq:eVmax} can be considered as an upper limit of the heat source for the cap surface. \\
\noindent As argued above, no atmosphere can exist within the IAR. Therefore, the total kinetic energy of the ultra-relativistic charges is assumed to be released in the surface layers of the real polar cap.\\
\noindent As the penetration depth $d_{PD}$ within the surface layers of the polar cap $\rho\sim 10^6$ g cm$^{-3}$ is exceedingly small (see Fig.~\ref{fig:pendepth}), the total released heat is immediately radiated away.\\
\noindent The backflowing charges that hit the real polar cap surface are perhaps not mono-energetic.   Based on observations of gamma-ray pulsars and simulations of pulsar magnetospheres (see \cite{CPS16,BKTHK18} and references therein), \cite{BPO19} suggest a power-law energy spectrum of the charges that form the return current given by

\beq
N(\gamma) = N_{0}\gamma^{\eta},
\label{eq:power-law}
\eeq  

\noindent where $N$ is the number of electrons for $\gamma > 1$, $N_{0}$ the normalization for $\gamma = 1$, $\eta$ is a free parameter. When $\eta$ is negative, as suggested by \cite{BPO19}, such a structured spectrum has a strong impact on the heating of the cap surface. Charges with lower energies are more numerous as their number decreases strongly with increasing $\gamma$. Clearly, the spectrum of the charges created in the IAR has another physical origin than that considered by \cite{BPO19}. However, we use the same power law ansatz to check whether the spectrum of the bombarding charges is more or less mono-energetic or how much it deviates from being mono-energetic to explain the observed $T_s$.\\

\begin{figure}
\begin{center}
\includegraphics[width=1.0\columnwidth]{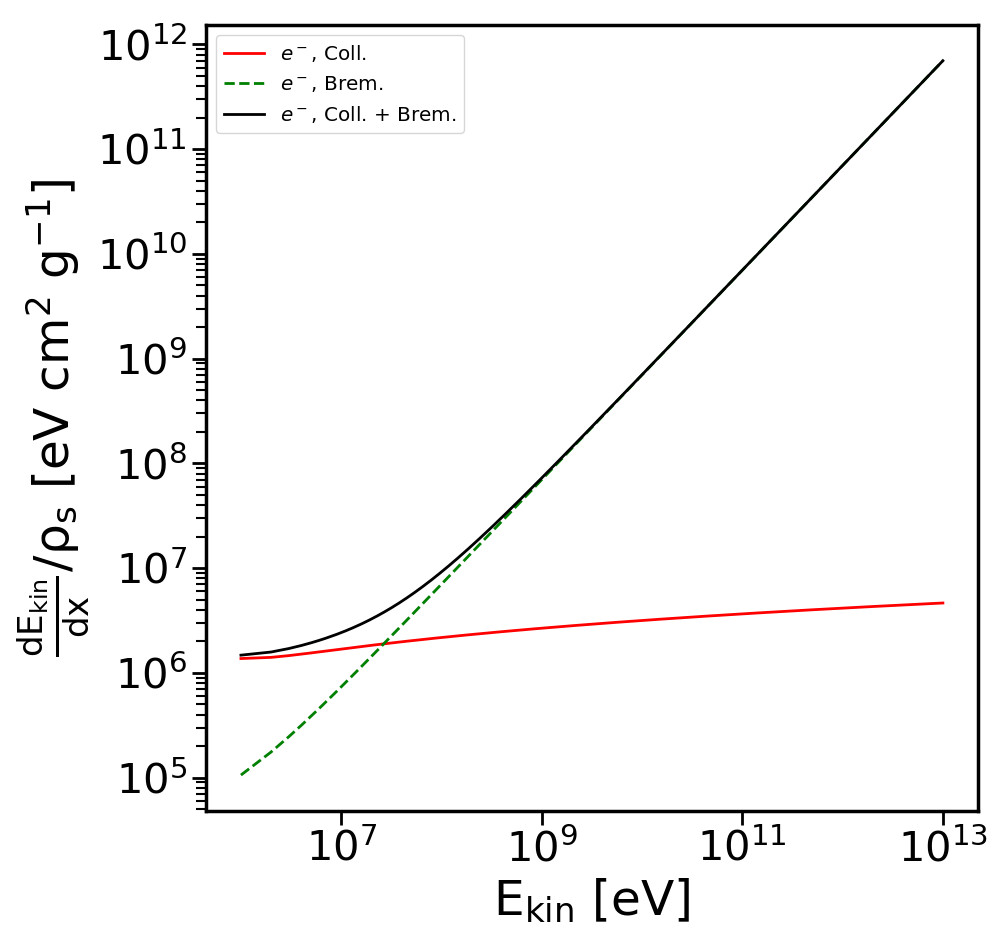}
\end{center}
\caption{Stopping power as a function of kinetic energy of electrons.}\label{fig:dE_dx}
\end{figure}

\subsection{Bethe-Bloch-Equation}

Electrons while passing through matter a distance $dx$ lose their kinetic energy $E_{kin}$ by collisions ${\frac{dE_{kin}}{dx}}_{coll}$ and Bremsstrahlung radiation ${\frac{dE_{kin}}{dx}}_{rad}$. For electrons passing through iron and having kinetic energy of $\approx 2.7 \times 10^{7}$ eV the ${\frac{dE_{kin}}{dx}}_{coll} = {\frac{dE_{kin}}{dx}}_{rad}$ \citep{LEO87}. Within the IAR, the kinetic energy of electrons bombarding the cap surface is certainly beyond this threshold. Fig. \ref{fig:dE_dx} shows the so-called stopping power $\frac{dE_{kin}}{dx}/\rs$ as a function of the electron kinetic energy. Clearly, above $\sim$$10^9$ eV, energy loss is dominated by Bremsstrahlung. Therefore, the only energy loss taken into account further on is due to that radiation. Hence, the Bethe-Bloch-Equation to be solved is:

\begin{eqnarray}
\frac{dE_{kin}}{dx} \equiv {\frac{dE_{kin}}{dx}}_{rad} &=& 4 N E_s Z^2 r_e^2 \alpha \cdot \\ \nonumber
& & \left[ \ln(183 Z^{-1/3}) + 1/18 - f(Z) \right]\,,
\label{eq:BBeq}
\end{eqnarray}

\noindent where $N = \rs N_{A} / A$ is the  particle number, $E_s$ is  the sum of kinetic and rest energy of an incident electron, $N_{A}$ is the Avogadro number, $r_{e}$ is classical electron radius, $\alpha \cong 1/137$  is the fine structure constant, while $f(Z)$ is

\begin{eqnarray}
  f(Z) &=& a^2 [(1 + a^2)^{-1} + 0.20206 - 0.0369a^2  \\ \nonumber
    & + & 0.0083a^4 - 0.002a^6]\,,
 \label{eq:f(z)}   
\end{eqnarray}

\noindent with $a = \alpha Z$ \citep{DBM54}. The amount of kinetic energy loss determines the  surface temperature $T_s$ of the real polar cap.

\subsection{Differential electron current}

In order to calculate the heat input from backflowing electrons at the bottom of the IAR, an estimate of the differential electron current is necessary. It is the magnitude of the current per electron kinetic energy , i.e. in units of $N$/s/eV.\\  
\noindent First, the electron current ($N$/s) has to be calculated. Assuming that it scales according to the power law of Eq. \ref{eq:power-law} 

\beq
j(\gamma) = j_0 \gamma^{\eta},
\label{eq:j_g}
\eeq

\noindent where $j=\frac{\partial{N}}{\partial t}$, $j_0$ represents then the current for $\gamma = 1$, and $j(\gamma)$ is the current for $\gamma > 1$. \cite{RS75} calculated the so-called \textit{maximum net charged particle flux} $\dot{N}_{max}$ which is the primary charge current accelerated in the IAR above the polar cap

\beq
	\dot{N}_{max} \approx S \frac{\vec{\Omega} \cdot \vec{B_s}}{2 \pi e}\,.
\label{eq:Nmaxdot}
\eeq

\noindent We identify, in variance to \cite{RS75},  $S=\pi R_{pc}^2$ not with the conventional but with the real polar cap surface and assume that $j_0 = \dot{N}_{max}$. By replacing in Eq. \ref{eq:j_g} the $\gamma$ factor by $E_{kin} = m_e c^2 (\gamma - 1)$ we obtain the differential electron current  ($j_0$ divided by $E_{kin}$). Hence, it is given by

\beq
j(B_s, E_{kin}) = S \frac{\vec{\Omega} \cdot \vec{B_s}}{2 \pi e E_{kin}} \left( \frac{E_{kin}}{m_e c^2} + 1 \right)^{\eta}\,.
\label{eq:e_spectrum}
\eeq  

\begin{figure*}
\centering
\begin{subfigure}[b]{0.45\textwidth}
   \includegraphics[width=1.0\columnwidth]{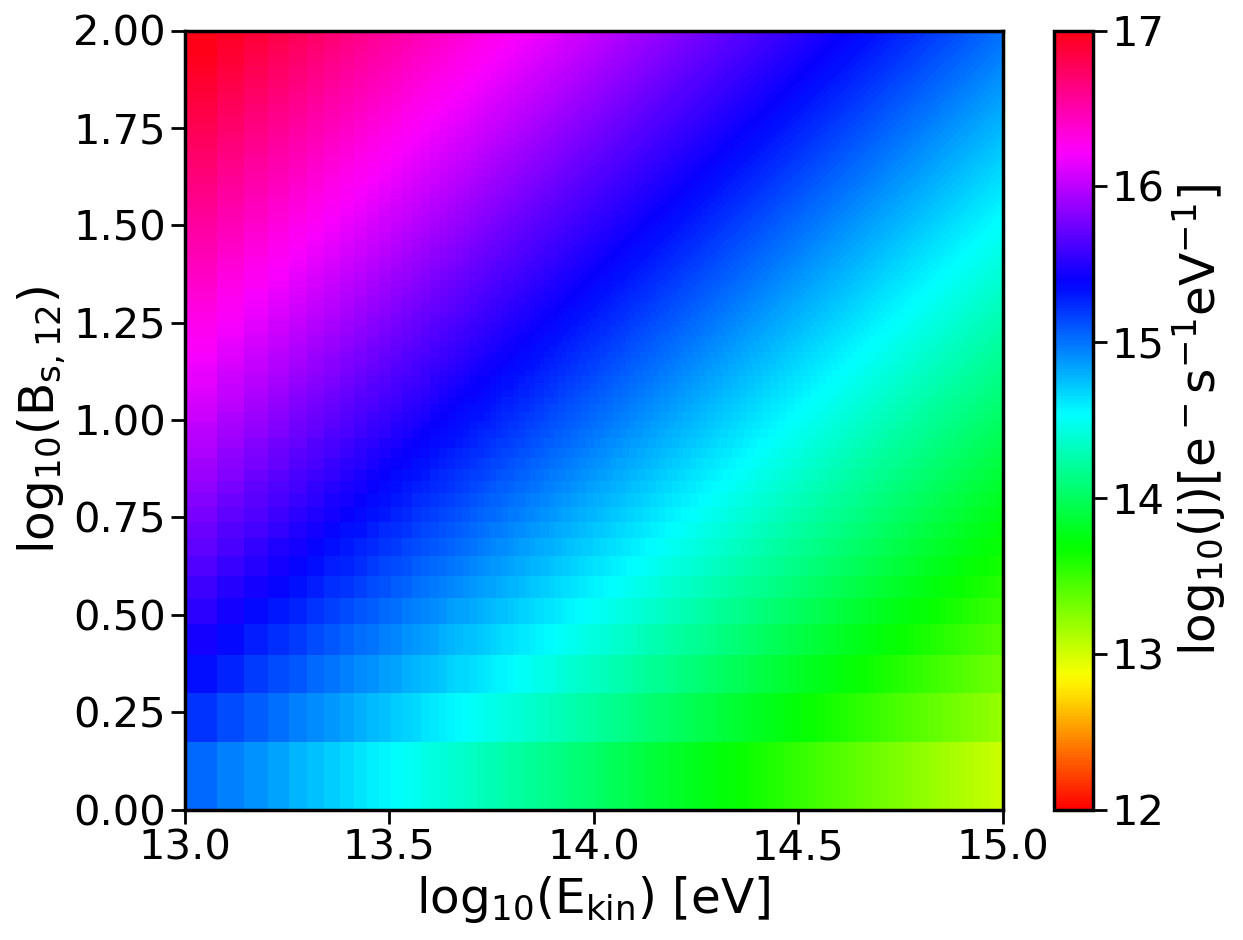}
   \caption{$\eta = -0.01$, P = 0.5}
   \label{fig:e_spectrum_m0_01_P0_5}
\end{subfigure}
~
\begin{subfigure}[b]{0.45\textwidth}
   \includegraphics[width=1.0\columnwidth]{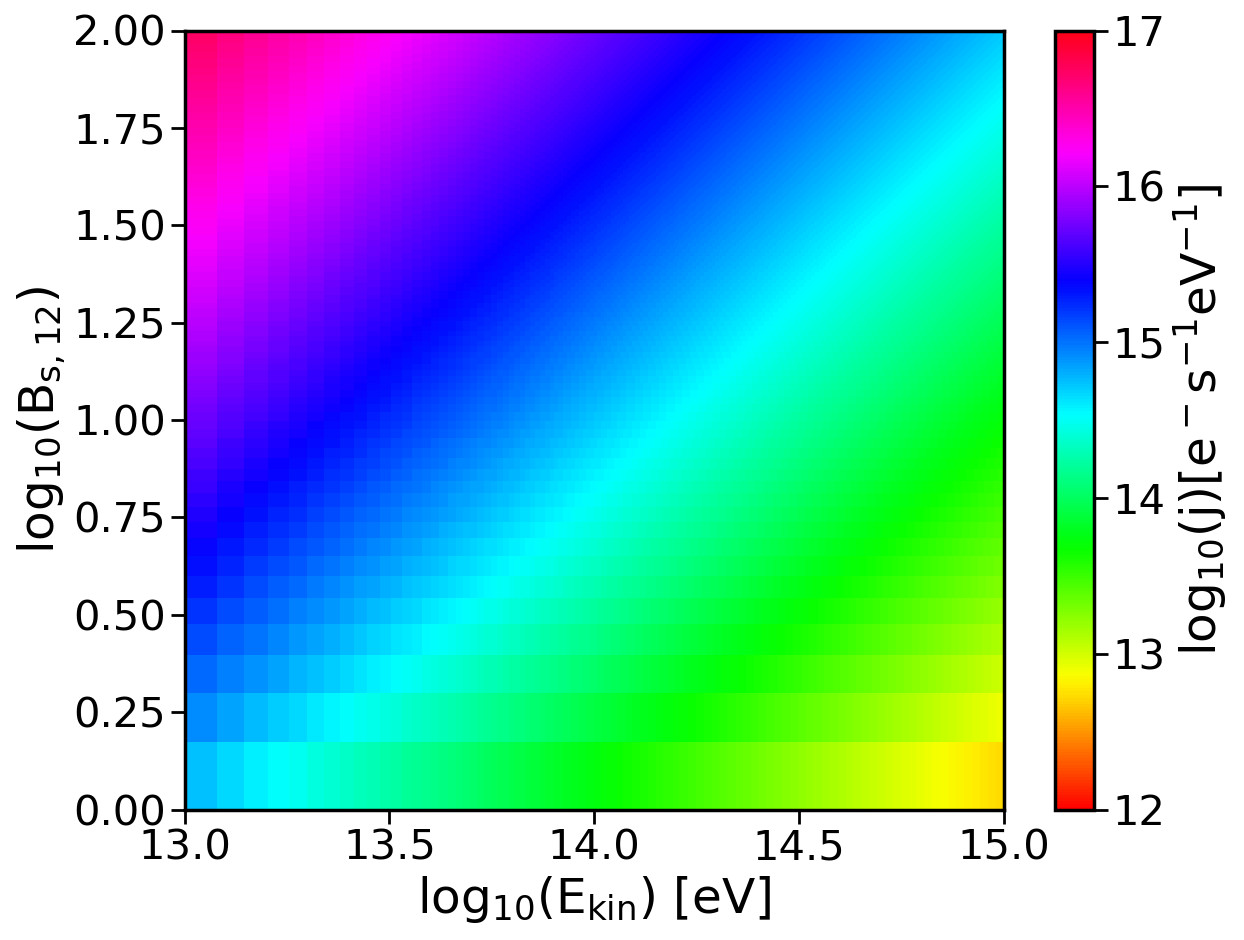}
   \caption{$\eta = -0.01$, P = 1.0}
   \label{fig:e_spectrum_m0_01_P1_0}
\end{subfigure}
~
\begin{subfigure}[b]{0.45\textwidth}
   \includegraphics[width=1.0\columnwidth]{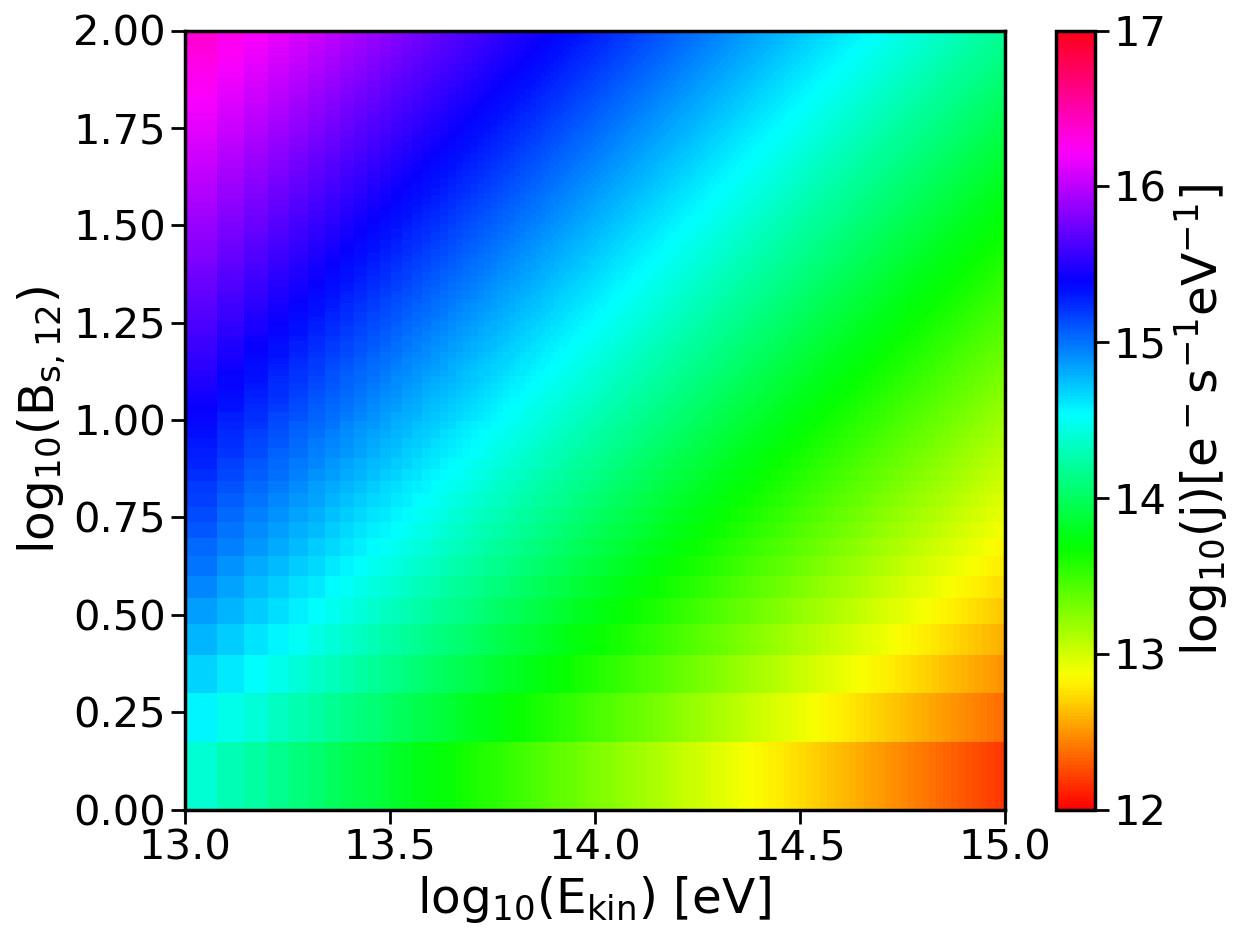}
   \caption{$\eta = -0.1$, P = 0.5}
   \label{fig:e_spectrum_m0_1_P0_5}
\end{subfigure}
~
\begin{subfigure}[b]{0.45\textwidth}
   \includegraphics[width=1.0\columnwidth]{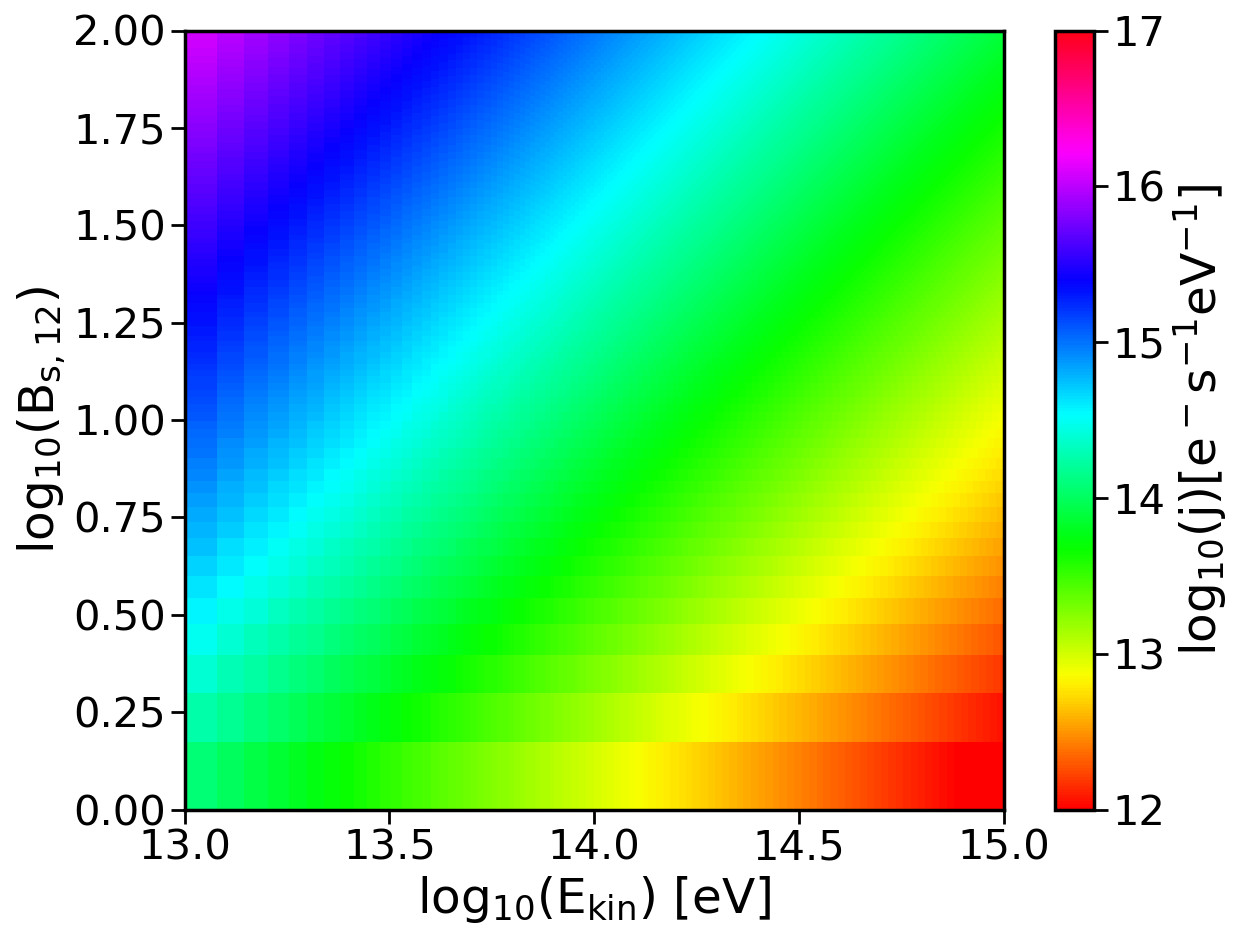}
   \caption{$\eta = -0.1$, P = 1.0}
   \label{fig:e_spectrum_m0_1_P1_0}
\end{subfigure}
\caption{Differential electron current spectra $j(B_s, E_{kin})$ for different $P$ and $\eta$.}
\label{fig:e_spectra}
\end{figure*}

\noindent The $j(B_s,E)$ spectrum is shown in Fig.\ref{fig:e_spectra} for $P = \{0.5 \mathrm{[s]}, 1.0\mathrm{[s]}\}$ and $\eta = \{-0.01, -0.1\}$. It is a relation between the magnetic field $B_{s,12}$ and electron kinetic energy. A variation of the parameter $\eta$ has a stronger impact on the current spectrum magnitude than a variation of the pulsar's rotational period $P$. 

\subsection{Heating}

Assume that all of the kinetic electron energy dissipated, while stopping down within real polar cap, is transferred into heat. By use of $j(B_s,E_{kin})$, energy loss $\frac{dE_{kin}}{dx}$, and the penetration depth $d_{PD}$ given by

\beq
d_{PD} = \int_{E_{electrons}}^{0} \left(\frac{dE_{kin}}{dx}\right)^{-1} \ dE\,,
\label{eq:dpd}
\eeq

\noindent we find the differential heating rate $Q'_{\mathrm{IN}}$. It is a measure how much heat is released within the real polar cap matter by the electrons per unit time and per electron kinetic energy:

\beq
Q'_{IN} = \frac{dE}{dx} d_{PD} j(B_s, E_{kin}),
\label{eq:heat}
\eeq

\begin{figure*}
\centering
\begin{subfigure}[b]{0.45\textwidth}
   \includegraphics[width=1.0\columnwidth]{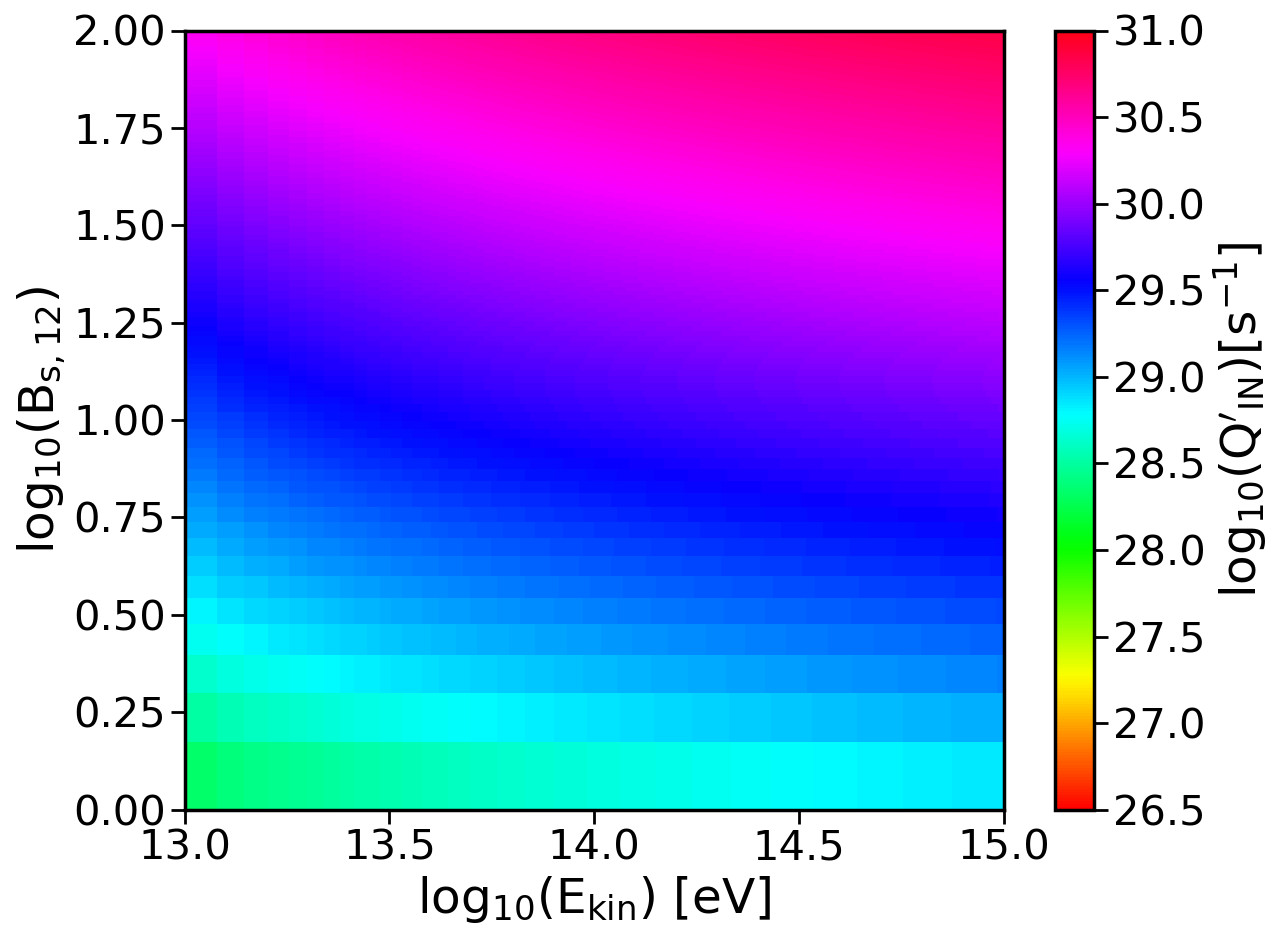}
   \caption{$\eta = -0.01$, P = 0.5}
   \label{fig:Q_spectrum_m0_01_P0_5}
\end{subfigure}
~
\begin{subfigure}[b]{0.45\textwidth}
   \includegraphics[width=1.0\columnwidth]{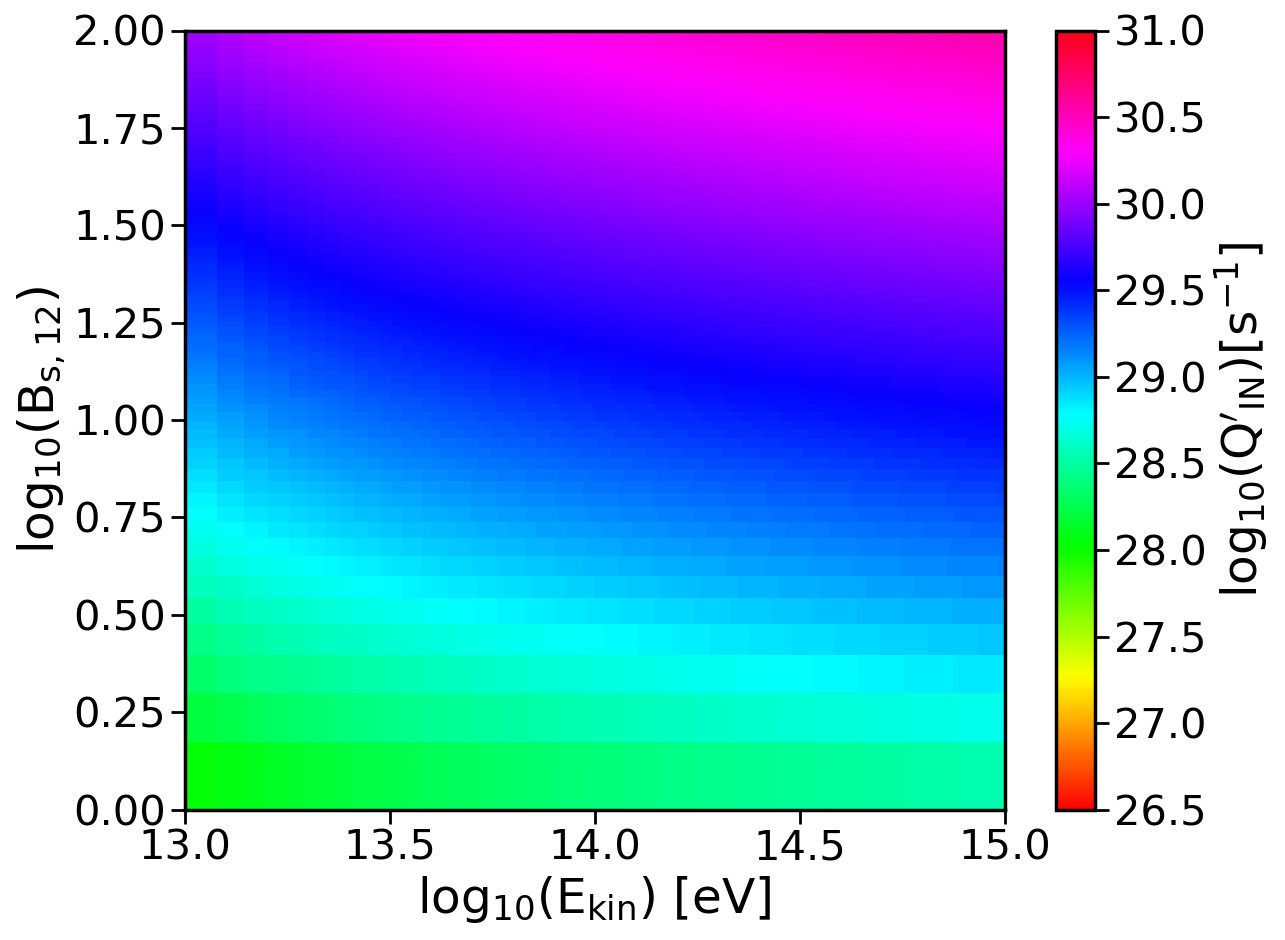}
   \caption{$\eta = -0.01$, P = 1.0}
   \label{fig:Q_spectrum_m0_01_P1_0}
\end{subfigure}
~
\begin{subfigure}[b]{0.45\textwidth}
   \includegraphics[width=1.0\columnwidth]{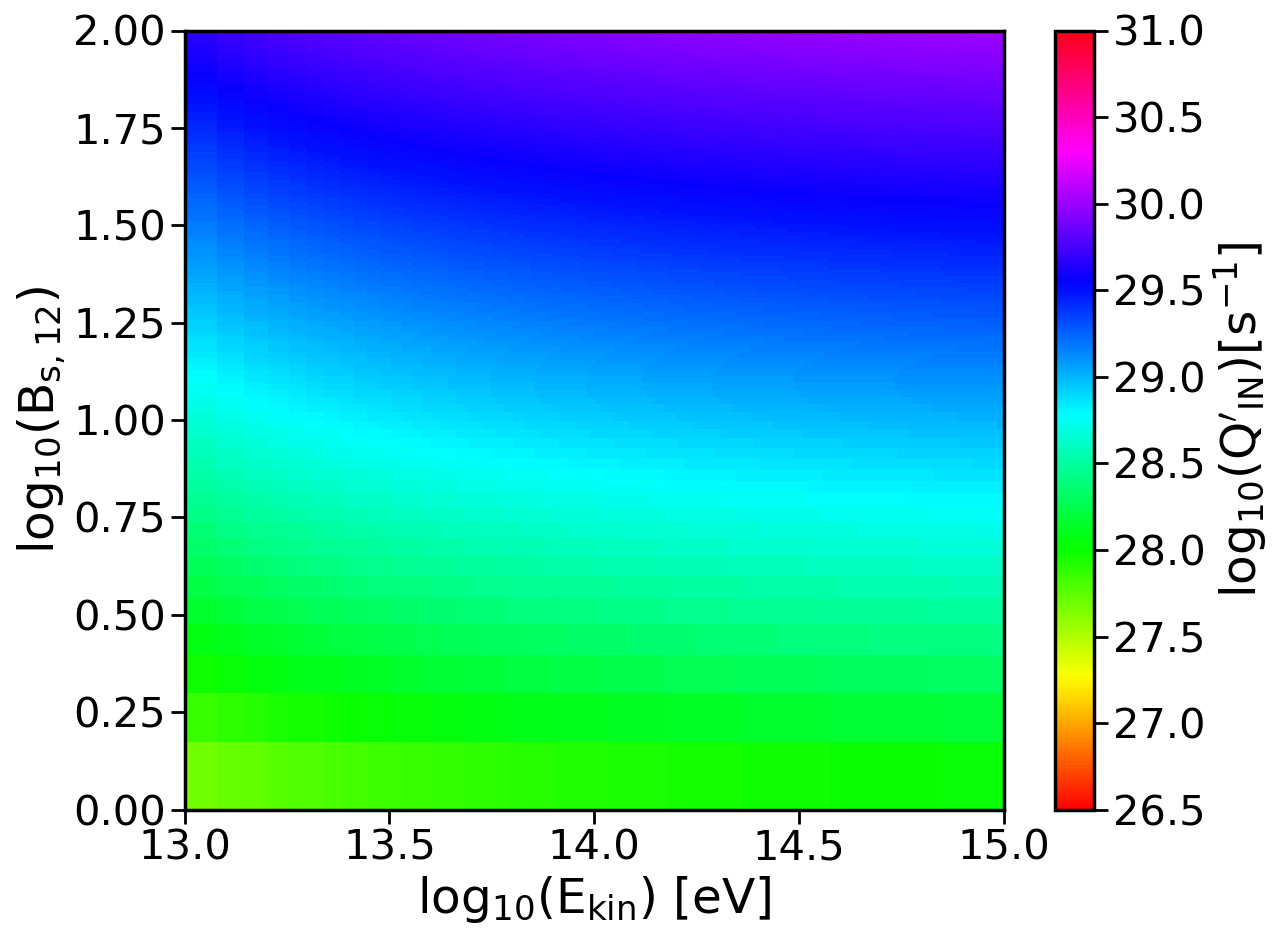}
   \caption{$\eta = -0.1$, P = 0.5}
   \label{fig:Q_spectrum_m0_1_P0_5}
\end{subfigure}
~
\begin{subfigure}[b]{0.45\textwidth}
   \includegraphics[width=1.0\columnwidth]{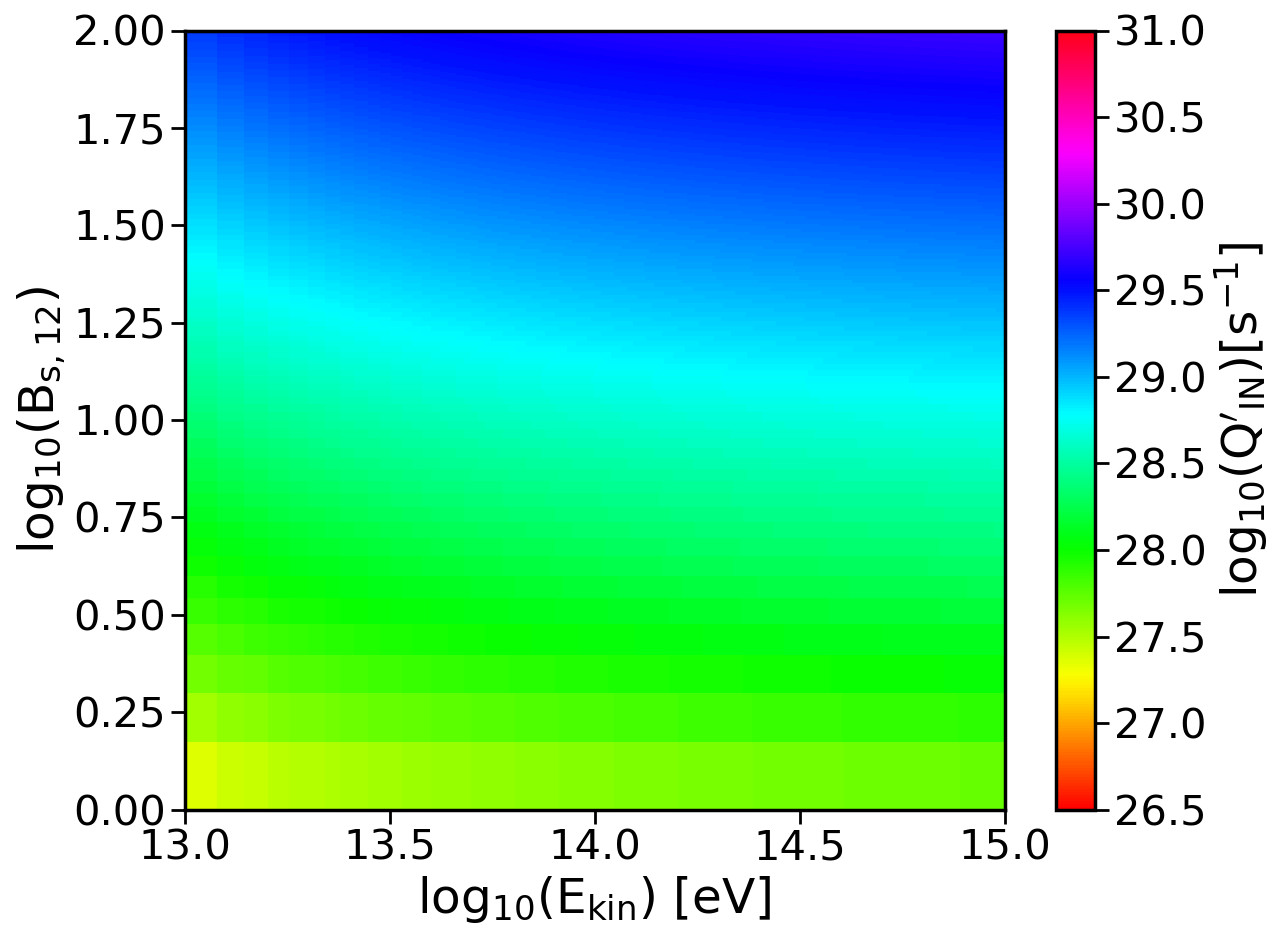}
   \caption{$\eta = -0.1$, P = 1.0}
   \label{fig:Q_spectrum_m0_1_P1_0}
\end{subfigure}

\caption{Differential heating rate spectra $Q'_{IN}$ for different $P$ and $\eta$.}
\label{fig:Q_spectra}
\end{figure*}

\noindent The differential heating rate spectra are shown in Fig. \ref{fig:Q_spectra}. As expected, $Q'_{IN}$ depends strongly on the power law parameter $\eta$.  

\noindent The heating rate $Q_{IN}$ is calculated by integrating $Q'_{IN}$ over the whole kinetic energy range of the electrons

\beq
Q_{IN} = \int Q'_{IN} \ dE\,.
\eeq

\noindent To calculate the temperature of the NS cap, assume that the $Q_{IN}$ is immediately radiated away from the very shallow polar cap surface layer into the space above it. This is justified by the smallness of $d_{PD}$ even for the highest kinetic energies (see Fig.~\ref{fig:pendepth}). A thermal balance will be established and maintained    

\beq
  Q_{IN} = Q_{OUT} = \sigma_{SB} \epsilon_{\mathrm{Fe}}ST_s^4,
\eeq

\noindent where $\sigma_{SB}$ is the Stefan-Boltzmann constant $3.54 \times 10^{-7}$ $\mathrm{eV \ cm^{-2} s^{-1} K^{-4}}$, $\epsilon_{\mathrm{Fe}}$ is the iron thermal emittance; assuming here $\epsilon_{\mathrm{Fe}}=0.6$. The temperature $T_s$ is then

\begin{eqnarray}
T_s = \left[ \frac{1}{2 \pi e \sigma_{SB} \epsilon_{\mathrm{Fe}}} \int \vec{\Omega} \cdot \vec{B_s} \frac{dE_{kin}}{dx} \frac{d_{PD}}{E_{kin}} \left(\frac{E_{kin}}{m_{e}c^2} + 1\right)^{\eta} dE\right]^{1/4}.
\label{eq:NS_T}
\end{eqnarray}

\noindent  Note that Eq. \ref{eq:NS_T} does not depend on target material density, since $\frac{dE_{kin}}{dx} \sim \rs$, while $d_{PD} \sim 1/\rs$.  The surface temperature also does not depend on the real polar cap area as $Q_{IN}\sim  j(B_s, E_{kin})\sim S$.

\section{Model validation}

\begin{figure*}
\centering
\begin{subfigure}[b]{0.45\textwidth}
	\includegraphics[width=1.0\columnwidth]{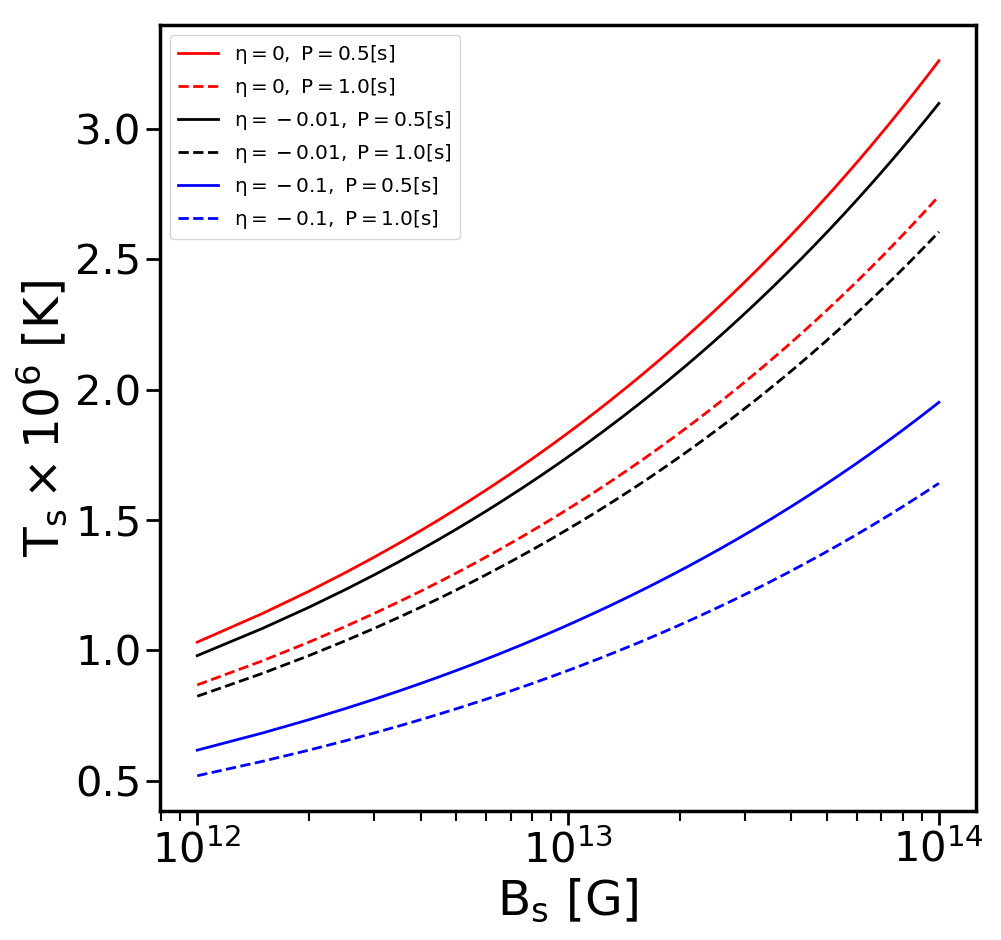}
	\caption{Model prediction}
	\label{fig:T_diff_eta_and_P}
\end{subfigure}
~
\begin{subfigure}[b]{0.45\textwidth}
	\includegraphics[width=1.0\columnwidth]{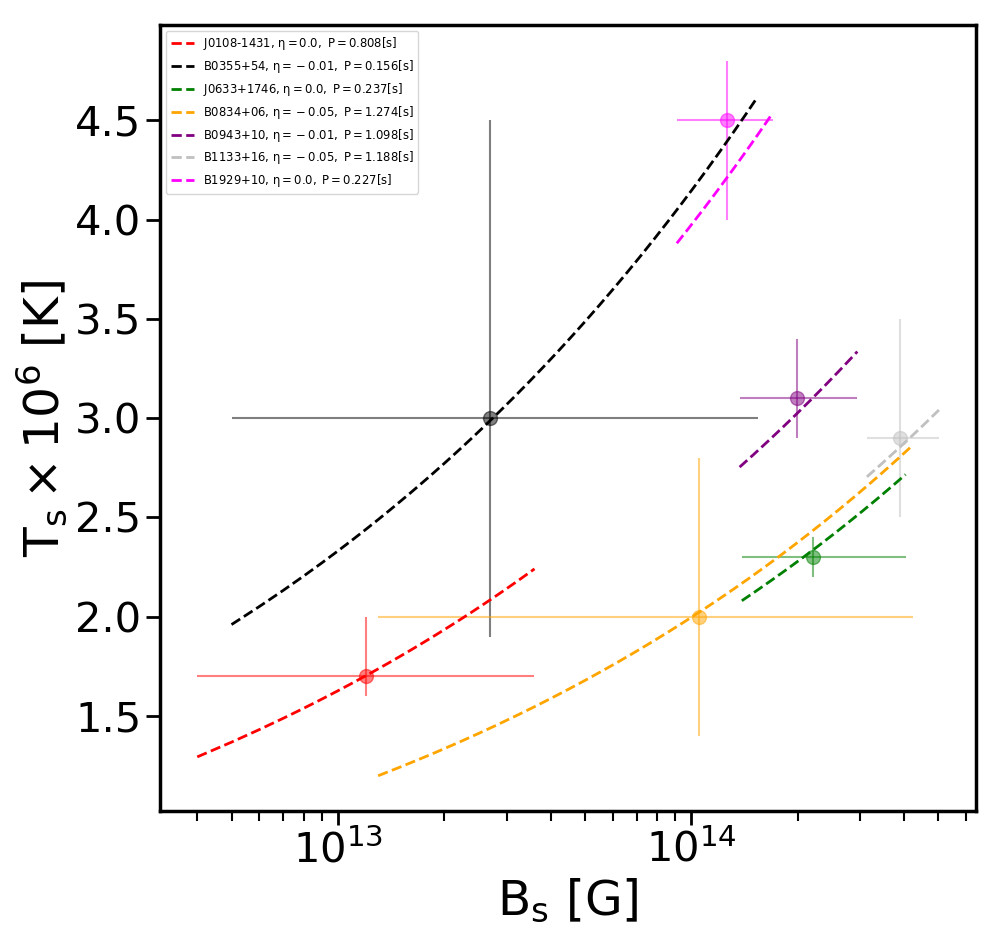}
	\caption{Model fit}
	\label{fig:T_diff_eta_and_P_FIT}
\end{subfigure}
\caption{Surface temperature of the real polar cap of radio pulsars as function of $B_s$ and $P$.}
\label{fig:T_eta_P}
\end{figure*}

\noindent Fig. \ref{fig:T_diff_eta_and_P} shows surface temperatures $T_s$ of the neutron star real polar cap as functions of the magnetic field strength $B_{s}$. The differential current spectrum of the bombarding ultra-relativistic electrons is described by Eq. \ref{eq:e_spectrum}. Consider $\eta=-0.1, -0.01$, and for comparison a mono-energetic spectrum $\eta=0$ as well as two rotational periods $P=0.5$ and $P=1.0$ s. Obviously, an almost mono-energetic spectrum of the primary charges returns the largest $T_s$. Its magnitude reaches $3\times10^6$ K for magnetic field strength $B_s$ of $\sim10^{14}$ G, a rotational period $P$ of 0.5 s, and $\eta=-0.01$. The rotational period has a smaller impact on the temperature than the slope of the spectrum.\\  
\noindent Our model has been validated by comparison with observational data of 7 pulsars (see Table \ref{tab:model_fit}). The validation has been performed by implementing a \textit{semi}-$\chi^2$ test procedure, i.e. each pulsar represents just one measuring point. Therefore, a typical $\chi^2$ fit reduces to one loop where $\eta$ was set as free parameter taken within a range from -0.5 to 0 in steps of 0.01. Hence, $\chi^2$ equals to the temperature difference of observed $T_s$ and the one calculated by Eq. \ref{eq:NS_T}. In Fig. \ref{fig:T_diff_eta_and_P_FIT} is shown the result of the validation procedure. The dashed lines represent the solutions of Eq. \ref{eq:NS_T} with the fitted $\eta$ - values. The results of the model calculations agrees well, within the error bars, with the observational values.

\begin{table}
	\centering
	\caption{Pulsar data used to fit the model parameter $\eta$ of the real polar cap surface temperature (Eq. \ref{eq:NS_T}); here $T_{s,6}$ is $T_s \times 10^6$ K and $B_{s,14}$ is $B_s \times 10^{14}$ G. Data originate from the following references (Ref.): (1) \citep{PAPMSK12} (2) \citep{MVKZCC07} (3) \citep{KPZR05} (4) \citep{GHMGZM08} (5) \citep{MTET13} (6) \citep{SGZHMGMX17} (7) \citep{MPG08}.}
	\label{tab:model_fit}
	\begin{tabular}{llcccc} 
		\hline
		Ref. & Name & P [s] & $T_{s,6}$ [K] & $B_{s,14}$ [G] & $\eta$ \\
		\hline
		1 & J0108-1431 & 0.808 & $1.7^{+0.3}_{-0.1}$ & $0.12^{+0.24}_{-0.08}$ & 0.0\\
		2 & B0355+54 & 0.156 & $3.0^{+1.5}_{-1.1}$ & $0.27^{+1.27}_{-0.22}$ & -0.01\\
		3 & J0633+1746 & 0.237 & $2.3^{+0.1}_{-0.1}$ & $2.21^{+1.83}_{-0.82}$ & 0.0\\
		4 & B0834+06 & 1.274 & $2.0^{+0.8}_{-0.6}$ & $1.05^{+3.19}_{-0.92}$ & -0.05\\
		5 & B0943+10 & 1.098 & $3.1^{+0.3}_{-0.2}$ & $1.99^{+0.96}_{-0.62}$ & -0.01\\
		6 & B1133+16 & 1.188 & $2.9^{+0.6}_{-0.4}$ & $3.9^{+1.12}_{-0.76}$ & -0.05\\
		7 & B1929+10 & 0.227 & $4.5^{+0.3}_{-0.5}$ & $1.26^{+0.44}_{-0.35}$ & 0.0\\
		\hline
	\end{tabular}
\end{table}

\noindent Some remarks about the application of these $T_s$-estimates on millisecond pulsars (MSPs). Recently, detailed NICER observation of PSR J0030+0451 were published by \cite{RWBRLGABBCGHHLMS19}, \cite{BWHRABGRGHC19} and \cite{2019ApJ...887L..24M}. These observations ($P\approx 4.87$ ms $\dot{P}\approx 1.02\times 10^{-20}$) return an estimate of the dipolar surface field strength $B_d\approx 2.3\times 10^8$ G and $R_{dip}\approx 2\times 10^5$ cm. The blackbody fit of the thermal X-ray spectrum assumes the existence of a hydrogen atmosphere, and indicates hot spots with $T_{s,6}\approx 1.3$. Radio emission of MSPs may originate either from an outer gap accelerator, or from the IAR above the polar cap (see e.g. discussion in \cite{JVHGSKCHFHLR14}). In the latter case, the requirement of potential gap formation demands a $B_s\gtrsim 5\times 10^{13}$ G (see \cite{ML07}, Fig.7). Inferred  $B_d$ close to the surface of MSPs
are 3 to 4 orders of magnitude weaker than in normal pulsars while charges at these regions experience an accelerating potential similar to that of normal pulsars. Therefore, flux conservation arguments result in a very small $R_{pc}\approx 430$ cm, smaller than any hot spot radius derived by \cite{RWBRLGABBCGHHLMS19} for different models of J0030+0451. The smallest $R_{pc}$ is observed for the normal PSR B1133+16 (see Tab.~\ref{tab:model_fit}, \cite{SGZHMGMX17}) where a blackbody fit results in $R_{pc}\approx 1400$ cm. To estimate radii as small as deduced here for MSP J0030+0451 is presently beyond the scope of observations.

\section{Conclusions}
The heat input into the surface layer and the resulting surface temperatures $T_s$ of the real polar cap were estimated and compared to available blackbody fits for seven radio pulsars. Our model assumptions were: 
\begin{enumerate}
 \item  real polar cap surface fields $B_s\sim 10^{13} \dots 10^{14}$ G estimated from flux conservation;
 \item  corresponding surface densities of the polar cap $\rho_s \sim 10^5 \ldots 10^6$ g cm$^{-3}$;
 \item  the polar cap surface consists of fully ionized solidified iron; 
 \item  in the IAR above the polar cap there was no hydrogen atmosphere;
 \item  the heat input was calculated  from the maximum $E_{kin}$ that primary particles acquire in the strong electric field prevalent within the IAR;
 \item although the bombardment of the cap surface is performed both by primary and the secondary charges created in pair cascades within the IAR, the maximum $E_{kin}$ is a reliable (upper) measure for the transfer of kinetic energy into heat and determines the total energy budget;
 \item a power law energy spectrum for the current of bombarding charges with a free exponent $\eta$ was assumed;
 \item due to the shallow penetration depth of bombarding charges into the cap surface, all the released kinetic energy was immediately re-radiated;
 \item a thermal balance was established and maintained at the real polar cap surface over the active life time of radio pulsars. 
\end{enumerate}
\noindent The following results were obtained:
\begin{enumerate}
\item $T_s$ were calculated as functions of $B_s$ and $P$;
\item these $T_s$ were compared with ``observed'' $T_s$ of seven radio pulsars for which both $B_s$ and $P$ were estimated and known, respectively;
\item since the accelerating electric field increases with $B_s$ and decreases with increasing $P$, $T_s$ was largest for the strongest $B_s$ and the most rapid rotation;
\item the smaller $\mid\eta\mid$ the higher $T_s$ for given values of $B_s$ and $P$;
\item a \textit{semi}-$\chi^2$ test by use of the observed pulsar parameters reveals that the spectrum of the bombarding charges was almost mono-energetic;
\item a relatively good agreement of model and observations was concluded.
\end{enumerate}
\noindent Clearly, taking the maximum kinetic energy of the charges we consider only an upper limit for the heat release in the cap surface. The overall heating process suffers from the intrinsic intermittency of pair cascades; it is not a continuous heating. The average $T_s$ (averaged over many discharging cycles) will be lower than those calculated when starting from the maximum kinetic energy. A quantitative estimate how much the real $T_s$ is smaller than that calculated by use of the maximum $E_{kin}$ is beyond the scope of this work. \cite{T10} and \cite{TA13} performed detailed simulations of pair cascade formation in the polar cap region. The cascade repetition rate could not yet be inferred directly from simulations, however, the reduction of heating efficiency might be significant (Timokhin 2020, private communication). This reduction could be mitigated somewhat. When comparing the cooling and heating time scales at the strongly magnetized cap surface for a discharge cycle (few $\mu$s) one finds $\tau_{heat}/\tau_{cool}<1$ (see Appendix of \cite{GMG03}), which makes the cooling in phases when the bombardment ceases less efficient.\\
\noindent Given the large error bars at the observed $T_s$ values, as shown in Fig. \ref{fig:T_diff_eta_and_P_FIT}, this good agreement was certainly not a satisfying proof. However, it indicates that our model, naked surface at the bottom of the IAR and $B_{s,12}\sim 100$, reflects the physics of the real polar cap heating of radio pulsars quite well.\\

\section*{Acknowledgments}
We gratefully acknowledge  G. Melikidze, A. Timokhin, and R. G. Bryant for enlightening discussions and a critical reading of the manuscript.

\bibliography{pulsars}
\end{document}